\documentstyle[12pt]{article}
\newcommand{\be}{\begin{eqnarray}}
\newcommand{\ee}{\end{eqnarray}}
\newcommand{\ba}{\begin{array}}
\newcommand{\ea}{\end{array}}

\newcommand{\beq}{\begin{equation}}
\newcommand{\eeq}{\end{equation}}
\newcommand{\fpi}{F_{\pi}}

\newcommand{\kstar}{K_{0}^{*}}

\begin{document}
\thispagestyle{empty}
\begin{center}
\section*{Scalar mesons in the Nambu--Jona-Lasinio model
with the 't Hooft interaction.}

\vskip0.5truecm
M. K. Volkov,  M. Nagy{\footnote{Permanent Address:
Institute of Physics, Slovak Academy of Sciences,
842 28 Bratislava, Slovakia}} , V. L. Yudichev
\end{center}
\date{}

\begin{center}
{\it
Bogoliubov Laboratory of Theoretical Physics,
Joint Institute for Nuclear Research, 141980 Dubna, Russian Federation
}
\end{center}

\begin{abstract}
\large
We calculate the mass spectra of the pseudoscalar and scalar meson
nonets in the Nambu--Jona-Lasinio model  with the
't Hooft interaction. We obtain satisfactory result for
the pseudoscalar mesons. For the scalar mesons, the 't Hooft
interaction somewhat increases the values of the masses. However,
it is not sufficient to explain the whole scalar mass spectrum.
The situation could  be
improved for the $\sigma$ and $f_0$ mesons through mixing with
the glueball state. For the description of the masses of
$a_0$ and $\kstar$ mesons, it is necessary to involve the other
models. The strong decay widths of the scalar mesons are described.
\end{abstract}

\newpage
\setcounter{page}{1}
\section{ Introduction}
During the last years noticeable progress has been achieved in both
experimental and theoretical investigations of scalar mesons.
The low-mass sigma meson appeared in the last Review of Particle
Properties (1996) \cite{Rev_96}. A new theoretical analysis of
experimental data on the low-mass sigma meson has been completed in the
papers \cite{Ishi_96,Svec_92}. Scalar mesons
in the NJL model with the 't Hooft interaction were investigated in
\cite{Dmitr_96,Celen_93}. In the papers \cite{Celen_93} the
coupling of the ${\bar q}q$ states to the two-pion continuum and
a model of confinement were also included into consideration.

Here we continue these investigations in the framework of the standard
NJL model with the 't Hooft interaction, following
the papers~\cite{Dmitr_96}--\cite{Kleva_92} .

Our work is organized as follows. In Sec. 2, we describe
the characteristics of our NJL model with the 't Hooft interaction.
We define the main parameters of this model - the cut-off parameter
$\Lambda$ and the constituent u-quark mass $m_u$ using the
experimental values of the pion decay coupling constant $F_{\pi}$=93
MeV, the $\rho$-meson coupling constant $g_{\rho}=6.14$ $({g^2_{\rho}\over
4\pi} \approx 3)$, describing the decay $\rho\to 2\pi$, and the relation
$g_{\rho}={\sqrt 6}g_{\sigma}$. In Sec. 3, we describe the masses of the
isovector and strange mesons. In Sec. 4, we calculate the quark-loop
contributions to the masses of the $\eta$, $\eta'$, $\sigma$ and $f_0$
mesons. In Sec. 5, the numerical estimations of the quark-loop
contributions to the meson masses are carried out.
In Sec. 6 the strong decay widths of the scalar mesons
are estimated.
Sec. 7 contains discussion and the conclusion.

\section{ The NJL model with the 't Hooft interaction}

The U(3)$\times$U(3) version of the NJL model supplemented by the 't Hooft
interaction takes the form
\be
L& =& {\bar q}(i{\hat \partial} - m^0)q + {G\over 2}\sum_{i=0}^8 [({\bar q}
{\lambda}_i q)^2 +({\bar q}i{\gamma}_5{\lambda}_i q)^2] -\nonumber\\
&&- K \left\{ {\det}[{\bar q}(1+\gamma_5)q]+{\det}[{\bar q}(1-\gamma_5)q]
\right\}
\label{Ldet}
\ee
where $\lambda_i$ (i=1,...,8) are the Gell-Mann matrices and
$\lambda^0 = {\sqrt{2\over 3}}${\bf 1}, with {\bf 1} being the unit matrix;
$m^0$ is a current quark mass matrix with diagonal elements $m^0_u$,
$m^0_d$, $m^0_s$ $(m^0_u \approx m^0_d)$. The Lagrangian (\ref{Ldet})
can be rewritten in the form (see \cite{Kleva_92})
\be
&&L = {\bar q}(i{\hat \partial} - m^0)q + {1\over 2}\sum_{i=1}^9[G_i^{(-)}
({\bar q}{\tau}_i q)^2 +G_i^{(+)}({\bar q}i{\gamma}_5{\tau}_i q)^2] +
\nonumber \\
&&\qquad+ G^{(-)}_{us}({\bar q}{\lambda}_u q)({\bar q}{\lambda}_s q)  +
G_{us}^{(+)}({\bar q}i{\gamma}_5{\lambda}_u q)({\bar q}i
{\gamma}_5{\lambda}_s q)~
\label{LGus}
\ee
where
\be
&&{\tau}_i={\lambda}_i ~~~ (i=1,...,7),~~~\tau_8 = \lambda_u = ({\sqrt 2}
\lambda_0 + \lambda_8)/{\sqrt 3},\nonumber\\
&&\tau_9 = \lambda_s = (-\lambda_0 + {\sqrt 2}\lambda_8)/{\sqrt 3}, \nonumber \\
&&G_1^{(\pm)}=G_2^{(\pm)}=G_3^{(\pm)}= G \pm 4Km_sI_1(m_s), \nonumber \\
&&G_4^{(\pm)}=G_5^{(\pm)}=G_6^{(\pm)}=G_7^{(\pm)}= G \pm 4Km_uI_1(m_u),
\nonumber \\
&&G_u^{(\pm)}= G \mp 4Km_sI_1(m_s), ~~~
G_s^{(\pm)}= G, ~~~
G_{us}^{(\pm)}= \pm 4{\sqrt 2}Km_uI_1(m_u).
\label{DefG}
\ee
Here $m_u$ and $m_s$ are the constituent quark masses and
\be
I^{\Lambda}_n(m_i)={N_c\over (2\pi)^4}\int d^4_e k {\theta (\Lambda^2 -k^2)
\over (k^2 + m^2_i)^n} .
\label{DefI}
\ee
Here we have used the Euclidean space and the cut-off parameter $\Lambda$.

Let us define the parameters $m_u$ and $\Lambda$, following the papers
\cite{Volk_82,Volk_86,Ebert_94}. We shall use the four equations

1) the Goldberger-Treiman relation
\be
g_{\pi{\bar q}q}={m_u\over F_{\pi}},
\label{Gold}
\ee

2) the relation between $g_{\rho}$ and $g_{\sigma{\bar q}q}$
\cite{Volk_82}-\cite{Kikka_76}
\be
g_{\rho} = {\sqrt 6}g_{\sigma{\bar q}q} ,
\label{grho}
\ee

3) the relation between $g_{\sigma{\bar q}q}$ and $g_{\pi{\bar q}q}$,
which was obtained by taking into account the $\pi - a_1$ transitions
(see \cite{Volk_86,Ebert_94})
\be
g_{\pi{\bar q}q} = Z^{-1/2} g_{\sigma{\bar q}q} , ~~~
Z=\left(1 - {6m^2_u\over M^2_{a_1}} \right)^{-1} ,
\label{Z}
\ee
where $M_{a_1}$ is the $a_1$-meson mass;

4) the expression for $g_{\sigma{\bar q}q}$ through a logarithmically
divergent integral \cite{Volk_82,Volk_86}
\be
g_{\sigma{\bar q}q}^2 = [4I^{\Lambda}_2(m_u)]^{-1}.
\label{log}
\ee

>From equations (\ref{Gold})-(\ref{Z}) it is possible to express
the constituent u-quark mass through the observable values $F_{\pi}=93$ MeV,
$g_{\rho}\approx 6.14$ and $M_{a_1}$ = 1230 MeV
\be
m^2_u = {M^2_{a_1}\over 12}\left[1-{\sqrt{1-\left({2g_{\rho}F_{\pi}\over
M_{a_1}}\right)^2}}\right], ~~~ m_u=280~{\rm MeV}.
\label{M}
\ee
Then, from (\ref{grho}) and (\ref{log}) we define
\be
\Lambda = 1250~{\rm MeV}.
\label{Lambda}
\ee

\section{The masses of the iso\-vec\-tor and strange mesons}

Now let us first consider  bosonization of the diagonal parts
of the Lagrangian (\ref{LGus}) including the isovector and
strange mesons. In Sec. 4, we complete the bosonization of the last
part of the Lagrangian (\ref{LGus}) containing the nondiagonal
terms.

After renormalization of the meson fields we obtain
\cite{Volk_82,Volk_86}
\be
L(\pi, K, a_0, K^*_0) = -{g_{\pi}^2\over 2G_{\pi}}{\vec \pi}^2
- {g_K^2\over G_K}K^2 - {g_{a_0}^2\over 2G_{a_0}}{\vec a}_0^2
- {g_{K_0^*}^2\over G_{K_0^*}}{K_0^*}^2 - \nonumber \\
-i{\rm Tr}~\ln\left\{1 -{1\over i{\hat \partial}-M}\left[\sum_{i=1}^7
(g_{\phi_i}i\gamma_5{\lambda_i}{\phi_i} +g_{\sigma_i}{\lambda_i}
{\sigma_i}) \right]\right\},
\label{piKa}
\ee
where $\phi_i$ and $\sigma_i$ are the pseudoscalar and scalar fields,
respectively, ${\vec \pi}^2$ = ${\pi^0}^2$ + $2\pi^+\pi^-$, $K^2$ =
$K^0{\bar {K^0}}$ + $K^+K^-$, ${\vec a}_0^2$ = ${a_0^0}^2$ +
$2a_0^+a_0^-$, ${K_0^*}^2$ = ${\bar{K_0^*}}K_0^*$ + ${K_0^*}^+
{K_0^*}^-$
\be
&&G_{\pi}=G + 4Km_sI_1(m_s), \nonumber \\
&&G_K=G_{\pi} - 4K(m_sI_1(m_s) - m_uI_1(m_u)), \nonumber \\
&&G_{a_0}=G_{\pi} - 8Km_sI_1(m_s),   \\
&&G_{K_0^*}=G_{\pi} - 4K(m_sI_1(m_s) + m_uI_1(m_u)), \nonumber
\label{Gpi}
\ee
\be
&&g^2_{a_0}=[4I_2(m_u)]^{-1},~~~g^2_{K_0^*}=[4I_2(m_u,m_s)]^{-1}, \nonumber \\
&&I_2(m_u,m_s)=
{N_c\over (2\pi)^4}\int d^4_e k {\theta (\Lambda^2 -k^2)
\over (k^2 + m^2_u)(k^2 + m^2_s)}=\nonumber\\
&&={3\over (4\pi)^2(m_s^2-m_u^2)}\left[m_s^2\ln\left({\Lambda^2
\over m_s^2}+1 \right) - m_u^2\ln\left({\Lambda^2\over m_u^2}+1 \right) \right],
\label{ga_0}\\
&&g_{\pi}=Z^{1/2}_{\pi}g_{a_0},~~~g_K=Z^{1/2}_Kg_{K_0^*},~~~Z_{\pi}\approx
Z_K\approx 1.44 \quad .\nonumber
\ee
Then, in the one-loop approximation, the following expressions for the
meson masses are obtained \cite{Volk_86}
\be
&&M^2_{\pi}=g^2_{\pi}\left[{1\over G_{\pi}} - 8I_1(m_u)\right], \nonumber \\
&&M^2_K=g^2_K\left[{1\over G_K} - 4[I_1(m_u)+I_1(m_s)]\right]+Z(m_s-m_u)^2,
\nonumber \\
&&M^2_{a_0}=g^2_{a_0}\left[{1\over G_{a_0}} - 8I_1(m_u)\right] + 4m^2_u , \\
&&M^2_{K_0^*}=g^2_{K_0^*}\left[{1\over G_{K_0^*}} - 4[I_1(m_u)+
I_1(m_s)]\right]+ (m_u+m_s)^2 .   \nonumber
\label{Mpi}
\ee
The first relation is used to define the parameter $G_{\pi}$.
For the experimental values $M_{\pi^0}=135$ MeV we have
\be
G_{\pi} = 4.92~{\rm GeV}^{-2}.
\label{G_pi}
\ee

\section{ The masses of the $\eta$, $\eta'$, $\sigma$ and $f_0$ mesons}

The nondiagonal part of the Lagrangian (\ref{LGus}) has the form
\be
{\Delta}L&&= {1\over 2}\left\{G_u^{(+)}({\bar q}i\gamma_5\lambda_u q)^2 +
2G_{us}^{(+)}({\bar q}i\gamma_5\lambda_u q)({\bar q}i\gamma_5\lambda_s q)+
G_s^{(+)}({\bar q}i\gamma_5\lambda_s q)^2+ \right. \nonumber \\
&&+G_u^{(-)}({\bar q}\lambda_u q)^2 \left.
+ 2G_{us}^{(-)}({\bar q}\lambda_u q)({\bar q}\lambda_s q) +
G_s^{(-)}({\bar q}\lambda_s q)^2 \right\} =  \\
&&= ({\bar q}i\gamma_5\lambda_{\alpha} q)T^P_{\alpha\beta}
({\bar q}i\gamma_5\lambda_{\beta} q)+({\bar q}\lambda_{\alpha} q)
T^S_{\alpha\beta}({\bar q}\lambda_{\beta} q), ~~(\alpha=u,s)~~(\beta=u,s)
\nonumber
\label{Ldelta}
\ee
where
\be
&&G_u^{(+)}=G_{\pi}-8Km_sI_1(m_s),~~~G_u^{(-)}=G_{\pi}, \nonumber \\
&&G_s^{(\pm)}=G_{\pi}-4Km_sI_1(m_s),~~~G_{us}^{(\pm)}=\pm
4{\sqrt 2}Km_uI_1(m_u) ,
\label{GuGs}
\ee
\be
T^{P(S)}=\frac{1}{2}  \left(
\begin{array}{cc}
G_u^{(\pm)}    & G_{us}^{(\pm)} \\
G_{us}^{(\pm)}  & G_s^{(\pm)}
\end{array}
\right)
\label{Tps}
\ee
After bosonization we obtain
\be
{\Delta} {\bar L}= -{g_{\eta_{\alpha}}g_{\eta_{\beta}}\over 4}\eta_{\alpha}
(T^P)^{-1}_{\alpha\beta}\eta_{\beta}-
{g_{\sigma_{\alpha}}g_{\sigma_{\beta}}\over 4}\sigma_{\alpha}
(T^S)^{-1}_{\alpha\beta}\sigma_{\beta} - \nonumber \\
-i~{\rm Tr}\ln \left\{1 + {1\over i{\hat \partial} - M}[g_{\eta_{\alpha}}
i\gamma_5 \lambda_{\alpha}\eta_{\alpha} + g_{\sigma_{\alpha}}
\lambda_{\alpha}\sigma_{\alpha} ] \right\}
\label{Lbar}
\ee
where
\be
(T^{P(S)})^{-1}=\frac{2}{G_u^{(\pm)}G_s^{(\pm)}-(G_{us}^{(\pm)})^2}
\left(
\begin{array}{cc}
G_s^{(\pm)}      & -G_{us}^{(\pm)} \\
-G_{us}^{(\pm)}  & G_u^{(\pm)}
\end{array}
\right) ,
\label{Tps1}
\ee
\be
g_{\sigma_u}=g_{\sigma{\bar u}u},~~~g_{\sigma_s}=[4I_2^{\Lambda}(m_s)]^{-1/2},
~~~g_{\eta_u}=g_{\pi{\bar u}u},~~~g_{\eta_s}=Z^{1/2}g_{\sigma_s}.
\label{gsigmau}
\ee
>From the Lagrangian (\ref{Lbar}), in the one-loop approximation, the
following expressions for the mass terms are obtained
\be
{\Delta}{\bar L}^{(2)}&&=-{g_{\eta_{\alpha}}g_{\eta_{\beta}}\over
4}\eta_{\alpha}(T^P)^{-1}_{\alpha\beta}\eta_{\beta}-
{g_{\sigma_{\alpha}}g_{\sigma_{\beta}}\over 4}\sigma_{\alpha}
(T^S)^{-1}_{\alpha\beta}\sigma_{\beta} + \nonumber \\
&&+4I_1(m_u)(g^2_{\eta_u}\eta_u^2 + g^2_{\sigma_u}\sigma_u^2) +
4I_1(m_s)(g^2_{\eta_s}\eta_s^2 + g^2_{\sigma_s}\sigma_s^2)-\\
&&-2(m^2_u \sigma^2_u + m^2_s \sigma^2_s) =
-{1\over 2}\left\{ \eta_{\alpha}M^{P}_{\alpha\beta}\eta_{\beta}
+\sigma_{\alpha}M^{S}_{\alpha\beta}\sigma_{\beta} \right\} \nonumber
\label{deltaL2}
\ee
where
\be
&& M^P_{uu}= g_{\eta_u}^2\left({1\over 2}(T^P)^{-1}_{uu} - 8I_1(m_u) \right),
\nonumber \\
&& M^P_{ss}= g_{\eta_s}^2\left({1\over 2}(T^P)^{-1}_{ss}-8I_1(m_s) \right), \\
&& M^P_{us}= {1\over 2}g_{\eta_u}g_{\eta_s}(T^P)^{-1}_{us}, \nonumber
\label{Mpuu}
\ee
\be
&& M^S_{uu}= g_{\sigma_u}^2\left({1\over 2}(T^S)^{-1}_{uu} - 8I_1(m_u) \right)
+4m^2_u , \nonumber \\
&& M^S_{ss}= g_{\sigma_s}^2\left({1\over 2}(T^S)^{-1}_{ss}-8I_1(m_s)
\right) + 4m^2_s , \label{Msuu} \\
&& M^S_{us}= {1\over 2}g_{\sigma_u}g_{\sigma_s}(T^S)^{-1}_{us}. \nonumber
\ee
After diagonalization of the Lagrangian (22) we find masses
of  the pseudoscalar and scalar mesons $\eta$,
$\eta'$, $\sigma$ and $f_0$
\be
M^2_{(\eta,\eta')}={1\over 2}\left[ M^P_{ss} + M^P_{uu} \mp
{\sqrt {(M^P_{ss}-M^P_{uu})^2 + 4(M^P_{us})^2}} \right],   \\
M^2_{(\sigma,f_0)}={1\over 2}\left[ M^S_{ss} + M^S_{uu} \mp
{\sqrt {(M^S_{ss}-M^S_{uu})^2 + 4(M^S_{us})^2}} \right].
\label{Meta2}
\ee
Let us define the mixing angle for the pseudoscalar mesons
\be
&&\eta_s = \eta \cos{\bar \theta} + \eta' \sin{\bar \theta}, \nonumber \\
&&\eta_u =-\eta\sin{\bar \theta}+\eta'\cos{\bar \theta},~~~
{\bar \theta}=\theta - \theta_0
\label{mix}
\ee
where $\theta_0 \approx 35.3^{\circ}$ is the ideal mixing angle $({\rm ctg}~
\theta_0={\sqrt 2})$ and $\theta$ is the singlet-octet mixing angle
\be
{\rm tg}~ 2{\bar \theta} = {2M^P_{us}\over -M^P_{ss}+M^P_{uu}}.
\label{tg2}
\ee
For the scalar mesons we use the relations
\be
&&\sigma_u = \sigma \cos {\bar \phi} + f_0 \sin {\bar \phi}, \nonumber \\
&&\sigma_s = -\sigma \sin {\bar \phi} + f_0 \cos {\bar \phi},
~~~{\bar \phi} = \theta_0 - \phi
\label{sigmas}
\ee
where $\phi$ is the singlet-octet mixing angle and
\be
{\rm tg}~ 2{\bar \phi} = {2M^S_{us}\over M^S_{ss}-M^S_{uu}}.
\label{tg2s}
\ee

\section{ Numerical estimations of quark-loop
contributions to meson masses}

Using for the parameters $m_s$ and $K$ the values
\be
m_s = 425~{\rm MeV},~~~ K = 13.3~{\rm GeV}^{-5}
\label{msK}
\ee
we obtain the following estimations for the masses of the pseudoscalar
and scalar mesons
\be
&&M_{\pi} = 135~{\rm MeV},~~~M_K = 495~{\rm MeV}, \nonumber \\
&&M_{\eta} = 520~{\rm MeV},~~~M_{\eta'} = 1000~{\rm MeV}.
\label{piKeta}
\ee
\be
&&M_{\sigma} = 550~{\rm MeV},~~~M_{f_0} = 1130~{\rm MeV}, \nonumber \\
&&M_{a_0} = 810~{\rm MeV},~~~M_{K_0^*} = 960~{\rm MeV}, \\
&&{\theta} = -19^{\circ},~~~\phi = 24^{\circ}.  \nonumber
\label{sigfaK}
\ee
Note that the pion and kaon masses are the input parameters in our model.
The experimental data are \cite{Rev_96}
\be
&&M_{\pi^0}=134.9764\pm0.0006~{\rm MeV},~~~M_{\pi^{\pm}}=139.6~{\rm MeV}, \nonumber \\
&&M_{K^+}=493.677\pm 0.016~{\rm MeV},~~~M_{K^0}=497.672\pm 0.031~{\rm MeV},
\nonumber \\
&&M_{\eta}=547.45\pm 0.19~{\rm MeV},~~~M_{\eta'}=957.77\pm 0.14~{\rm MeV},
\nonumber\\
&&\theta\approx -20^{\circ}\quad\cite{GilKauff}.
\label{Mexp}
\ee
\be
&&M_{\sigma_0(400-1200)}=400 - 1200~{\rm MeV},~~~M_{f_0(980)}=980\pm 10~{\rm MeV},
\nonumber \\
&&M_{a_0}=983.5\pm 0.9~{\rm MeV},~~~M_{K_0^*}=1429\pm 6~{\rm MeV}.
\label{MexpfK}
\ee
Comparing the theoretical results with the experimental data we can
see that we have obtained satisfactory results for the pseudoscalar mesons
and for the octet-singlet mixing angle of the $(\eta \eta')$ mesons.
However, for the scalar mesons we have got the masses smaller than
the experimental data (except for the $f_0$ meson).

\section{Strong decays of the scalar mesons.}
Now, let us show what the information concerning the
strong decays of the scalar and pseudoscalar mesons
we can obtain from the Lagrangian~(\ref{piKa}) and~(\ref{Lbar}).
These Lagrangians allow us to get the following expressions for
the scalar-pseudoscalar meson vertices, describing the corresponding
strong decays of the scalar mesons
\be
g_{\sigma\pi\pi}&=& \frac{2m_u^2Z^{1/2}}{\fpi}\cos\bar\phi\nonumber\\
g_{f_0\pi\pi}&=& \frac{2m_u^2Z^{1/2}}{\fpi}\sin\bar\phi\nonumber\\
g_{a_0\eta\pi}&=& \frac{2m_u^2Z^{1/2}}{\fpi}\sin\bar\theta\\
g_{K_0^{*+}K^-\pi^0}&=& \frac{2m_u^2Z^{1/2}}{\fpi}\nonumber\\
g_{K_0^{*+}K^0\pi^-}&=& \frac{2\sqrt{2}m_u m_sZ^{1/2}}{\fpi}\nonumber
\ee
With these vertices one obtains the following decay widths
\be
\ba{lclclcl}
\Gamma_{\sigma\to\pi\pi}&\approx& 700~~{\rm MeV},&\quad&
	\Gamma_{f_0\to\pi\pi}&\approx& 20~~{\rm MeV},\\
\Gamma_{a_0\to\eta\pi}&\approx& 130~~{\rm MeV},&\quad&
	\Gamma_{\kstar\to K\pi}&\approx& 330~~{\rm MeV}.
\ea
\ee
These values are in qualitative agreement with the experimental
data
\be
\ba{lclclcl}
\Gamma_{\sigma\to\pi\pi}^{exp}&\sim& (600-1000)~~{\rm MeV},&\quad&
	\Gamma_{f_0\to\pi\pi}^{exp}&\sim& (30-78)~~{\rm MeV},\\
\Gamma_{a_0\to\eta\pi}^{exp}&\sim& (50-100)~~{\rm MeV},&\quad&
	\Gamma_{\kstar\to K\pi}^{exp}&\approx& (263\pm26\pm21)~~{\rm MeV}.
\ea
\ee

\section{ Conclusion}
Our calculations have shown that the 't Hooft interaction allows us  to describe the
masses of pseudoscalar meson's masses and their singlet-octet mixing angle in
satisfactory agreement with the experiment\footnote{
In \cite{Volk_82}--\cite{Volk_86} the pseudoscalar
$\eta$ and $\eta'$ meson's masses
have been described by means of introducing an additional isoscalar
quadratic term
into the meson Lagrangian,
connected with the gluon anomaly. There were obtained
results very close to our work.}.
 For the scalar meson masses
we have also obtained results more close to experimental data  than
in the NJL model without the 't Hooft interaction. However, the masses of
the $a_0$ and especially the $K_0^*$ mesons are
noticeably less then the experimental ones.

In order to get satisfactory result for the $f_0$ meson it is necessary to take
into account the mixing of the $f_0$ and $\sigma$ mesons with the glueball
state (see \cite{Kusaka_93}). The problem with
masses of $a_0$ and $K_0^*$ mesons could be solved in the framework
of the four-quark $(\bar q^2q^2)$ MIT-bag model \cite{MITbag}
or the $\bar KK$ molecule model \cite{KKmolec}.
However, it is interesting to note that in spite of
a very rough description of the scalar meson's masses our
model allows us to obtain a qualitatively true
picture of the strong decay widths of the scalar mesons.

\begin{center}
\section*{ Acknowledgments}
\end{center}

This work has been supported in part by the grant from RFFI
No. 98-02-16135 and the Heisenberg-Landau program, 1998.

\vskip2.0truecm



\begin{thebibliography}{99}
%
\bibitem{Rev_96} Review of Particle Properties, Phys. Rev.
{\bf D54} (1996) 1.
%
\bibitem{Ishi_96} S. Ishida et al. - Progr. Theor. Phys. {\bf 95}
(1996) 745.
%
\bibitem{Svec_92} M. Svec, A. de Lesquen, L. van Rossum - Phys. Rev.
{\bf D42} (1992) 949; hep-ph/9511205.
%
\bibitem{Dmitr_96} V. Dmitra\v sinovi\'c - Phys. Rev. {\bf C53} (1996) 1383.
%
\bibitem{Celen_93} L. S. Celenza, C. M. Shakin, J. Szweda - Int. J.
Mod. Phys. {\bf E2} (1993) 437; \\
L. S. Celenza, Xiang-Dong, C. M. Shakin - Phys. Rev. {\bf C56} (1997)
3326.
%
\bibitem{Kleva_92} S. P. Klevansky - Rev. Mod. Phys. {\bf 64} (1992) 649.
%
\bibitem{Volk_82} M. K. Volkov, D. Ebert - Sov. J. Nucl. Phys. {\bf 36}
(1982) 736; Z. Phys. {\bf C16} (1983) 205; \\
M. K. Volkov - Ann. Phys. {\bf 157} (1984) 282.
%
\bibitem{Volk_86} M. K. Volkov - Sov. J. Part. and Nuclei {\bf 17}
(1986) 186.
%
\bibitem{Ebert_94} D. Ebert, H. Reinhardt, M. K. Volkov - Progr. Part.
Nucl. Phys. {\bf 35} (1994) 1.
%
\bibitem{Kikka_76} H. Kikkawa - Progr. Theor. Phys. {\bf 56} (1976) 974.
%
\bibitem{GilKauff} F. J. Gilman and R. Kauffman, Phys. Rev. {\bf D 36}
(1987) 2761.
%
%
%
\bibitem{Kusaka_93} K. Kusaka, M. K. Volkov, W. Weise - Phys. Lett.
{\bf B302} (1993) 145.
%
\bibitem{MITbag} R. L. Jaffe, Phys. Rev. {\bf D15} (1977) 267,
{\bf D15} (1977) 281.
\bibitem{KKmolec} J. Weinstein and N. Isgur, Phys. Rev.
{\bf D41} (1990) 2236.
\end{thebibliography}
\end{document}